\documentclass[aps,prd,twocolumn]{revtex4}
\usepackage{graphicx}
\usepackage[switch,columnwise]{lineno}

\begin{document}

\title{Uranium and Plutonium antineutrino spectra taken by summation method used to fit experimentally measured reactor antineutrino spectrum}

\author{P.~Alzhev,$^{1,2}$ S.~Ingerman,$^{1}$ P.~Naumov,$^{2}$ V.~Sinev,$^{1,2}$ A.~Vlasenko$^{1,2}$}

\affiliation{$^{1}$ Institute for Nuclear Research of Russian Academy of Sciences, Moscow, Russia}
\affiliation{$^{2}$ National Research Nuclear University MEPHI, Moscow, Russia}

\begin{abstract}
Antineutrino spectra of fissile isotopes calculated using summation method being used to fit experimental spectra got in several experiments. Summation method was used with the possibility of changing individual spectrum shape of unknown fragments. Found spectra conform to all experiments and reproduce with the best accuracy measured inverse beta decay cross sections.
\end{abstract}
\maketitle

%\linenumbers

\section{Introduction}

For the purpose of nuclear reactor monitoring one needs to know exact antineutrino spectrum components. For the moment calculated spectra are not satisfied to the measured ones. They can’t predict measured reactor antineutrino spectrum shape and inverse beta-decay reaction cross section calculated with measured spectrum. Components of reactor antineutrino spectrum consist of the spectra produced by fission fragments of main fissile elements, they are $^{235}$U, $^{238}$U, $^{239}$Pu and $^{241}$Pu. The calculation of these spectra can be made by summing of individual antineutrino spectra from fissile isotope array of fission fragments. This is summation method, that is commonly used.

To calculate individual fragment spectrum, one should use a database for beta-decay isotopes. They contain information on beta-decay branches and their probabilities as well as fission yields, that is a probability to be born directly during the fission of one of fissile elements. There exist a number of databases like this, usually they belong to the most famous nuclear centres.

The most exact antineutrino spectra could be found if to describe experimentally measured antineutrino spectrum. Up today the are four known measured spectra from nuclear reactor, they are: Rovno spectrum done in form of a formulae valid from 2 up to 9 MeV \cite{rovno}, Double Chooz spectrum \cite{kerret} found from 1.8 up to 10 MeV and Daya Bay spectrum with the highest statistics laying in energy range 1.8 up to 8 MeV and the part of spectrum from 8 to 12 MeV \cite{an}. There was also published a paper from the RENO experiment with unfolded antineutrino spectrum \cite{yoon}.

If to fit experimental antineutrino spectrum with sum of calculated spectra one can find exact components belonging to spectra from $^{235}$U, $^{238}$U, $^{239}$Pu and $^{241}$Pu. The problem is to find the possibility to change calculated spectra for making a fit procedure. In this paper we describe the method of fitting experimental antineutrino spectrum with calculated ones using supposed beta-decay branches for unknown fission fragments nuclei.

\
\section{Beta-decay database}

To calculate antineutrino spectra we made our own database on fission fragments beta-decay properties basing on the known interactive database Livechart \cite{livechart}. Our database contain data on beta-decay branches for every nuclide and direct yields per fission of known heavy isotopes of uranium and plutonium.

While making the database we paid attention on the fact that all nuclides can be divided on three almost equal parts: surely known data, that were experimentally measured; partially known or estimated data, that couldn’t be measured but were calculated establishing on the similarity of nuclear properties of similar nuclei; totally unknown except of beta-decay energy Q and approximate half-life. The third part of nuclides placed far from the line of isotopes that are beta-stable. The first part is neighbouring to the beta-stable isotopes.

If to use hypothetical beta-decay branches for unknown nuclides in spectra calculation it may be possible to satisfy the known experimental antineutrino spectra. The procedure we have used in fitting experimental spectra is shown below.

\section{Measured experimentally reactor antineutrino spectra}

We regard four experimental antineutrino spectra with high statistics that were performed up today. All spectra were measured through inverse beta-decay reaction (IBD)

\begin{equation}
\bar{\nu}_{e} + p \rightarrow n + e^{+}.
\label{ibd}
\end{equation}

First one is the experiment done at the end of 80th years last century in former Soviet Union at Rovno \cite{rovno}. It accumulated 174 thousand neutrino events during three years data taking. In \cite{rovno} antineutrino spectrum was presented as a formula, what was useful to make estimations of neutrino yield in a detector placing at some distance from nuclear reactor. Later the spectrum was taken from the same data using the developed method of transformation a positron spectrum from inverse beta-decay (IBD) reaction (\ref{ibd}) to the antineutrino one through the function calculated from Monte Carlo simulation \cite{sinev}.

The second one was presented by Double Chooz experiment \cite{kerret}. The measured positron spectrum was transformed to the neutrino one by the same method as in case of Rovno. Its total statistics taken by the near detector is a little bit more than 200 thousand events.

The third spectrum was presented by the Daya Bay Collaboration\cite{an}. They used Single Value Decomposition (SVD) method to get antineutrino spectrum weighted with cross section. We transformed it to the antineutrino one. The spectrum includes several million neutrino events at near position of a detector. 

The last antineutrino spectrum was performed by the RENO Collaboration \cite{yoon}. They also used SVD method to obtain antineutrino spectrum weighted with cross section. There are about one million neutrino events in the spectrum.

\begin{figure}[ht]
\begin{center}
\includegraphics[width=86mm]{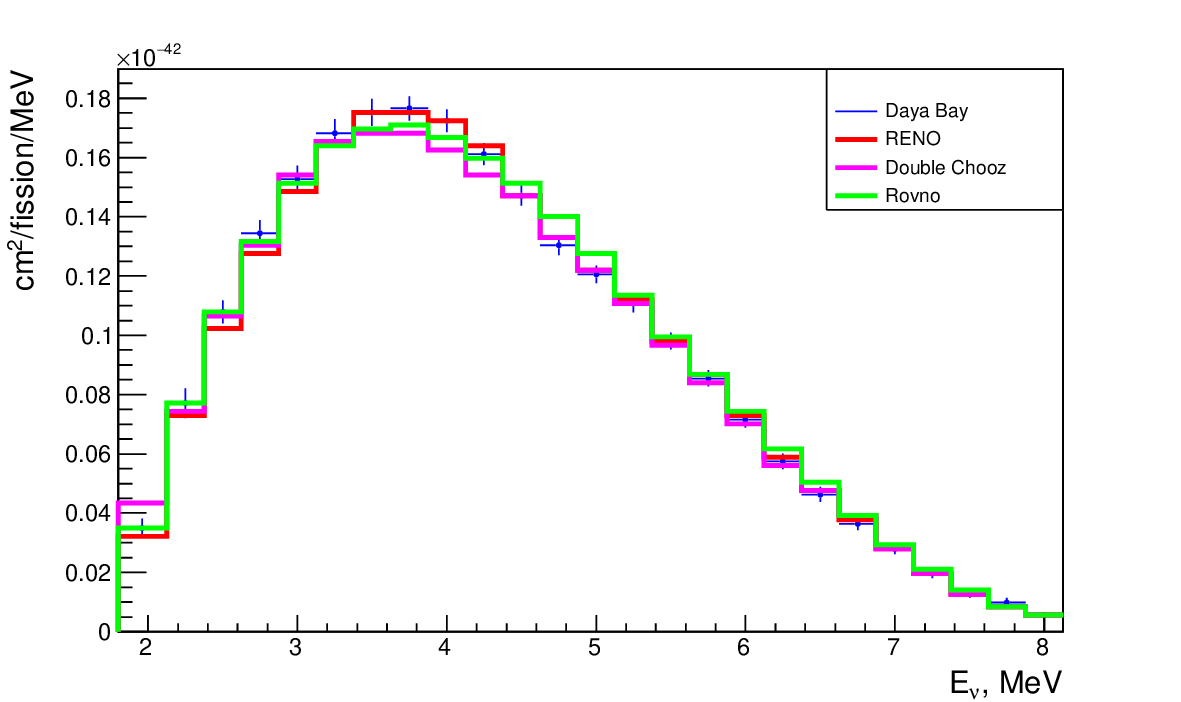}
\end{center}
\caption{\label{figone} Antineutrino spectra weighted with IBD cross section from experiments Rovno \cite{rovno}, Double Chooz \cite{kerret}, Daya Bay \cite{an} and RENO \cite{yoon}.}
\end{figure}

All mentioned above experimental antineutrino spectra are shown at Fig. \ref{figone}.

\section{Fitting of experimental antineutrino spectrum with the calculated ones}

\begin{figure}[ht]
\begin{center}
\includegraphics[width=86mm]{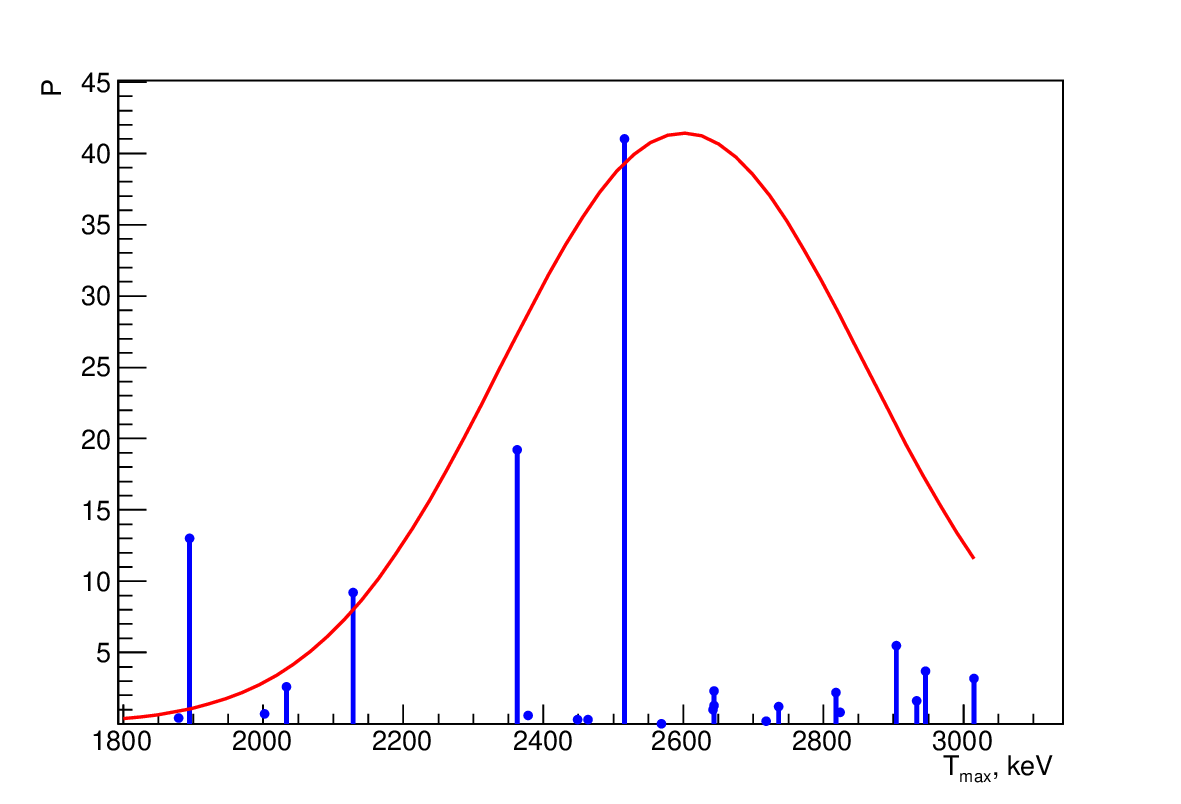}
\end{center}
\caption{\label{figtwo} Branch probabilities for $^{144}$Ba and proposed hypothetical branches, shown as gaussian. Gaussian probabilities were used for changing nuclide spectrum shape.}
\end{figure}

In 1981-1989 measurements of electron spectra from fission fragments of $^{235}$U, $^{239}$Pu and $^{241}$Pu they used a set of hypothetical beta-spectra to describe experimental spectrum. We use the calculation by summation method with variation of individual unknown spectra by changing their branch probabilities. At Fig. \ref{figtwo} one can see branch probabilities for nuclide $^{144}$Ba as an example. It is the nucleus that we mark as estimated. At the same figure we show our hypothetical distribution of branch probabilities for the nuclide $^{144}$Ba as a gaussian function. As one can see the gaussian distribution in common satisfy to the estimated distribution. 

Using hypothetical branches probabilities, we can change individual nuclide spectrum shape. This method allows to fit experimental antineutrino spectrum varying unknown spectra shapes. We do several calculations changing branch probability distribution for the chosen nuclide. Mean energy value $\mu$ of the distribution runs through the range from Q down to 2 MeV, while dispersion is $0.1\mu$. At each spectrum calculation we compose $\chi^{2}_{k}$ value and choose the minimal one.  Then we go to the next unknown nuclide and repeat this procedure. Used for minimisation $\chi^{2}_{k}$ shown in (\ref{chi2}).

\begin{eqnarray}
\chi^{2}_{k} = \sum_{DC}\frac{(y_{exp,i}-y_{calc,i})^2}{\Delta y_{i}^2}+\sum_{DB}\frac{(y_{exp,j}-y_{calc,j})^2}{\Delta y_{j}^2}+ \nonumber \\
\sum_{RENO}\frac{(y_{exp,l}-y_{calc,l})^2}{\Delta y_{l}^2}+\sum_{Rovno}\frac{(y_{exp,m}-y_{calc,m})^2}{\Delta y_{m}^2}+\label{R_CNO_EXP} \nonumber \\
\frac{(\sigma_{DC}-\sigma_{calc,DC})^2}{\Delta \sigma_{DC}^2}+\frac{(\sigma_{DB}-\sigma_{calc,DB})^2}{\Delta \sigma_{DB}^2} \nonumber \\
\frac{(\sigma_{RENO}-\sigma_{calc,RENO})^2}{\Delta \sigma_{RENO}^2}+\frac{(\sigma_{Bugey}-\sigma_{calc,Bugey})^2}{\Delta \sigma_{Bugey}^2},
\label{chi2}
\end{eqnarray}
where first term is for the Double Chooz spectrum, the second for the Daya Bay one, third for the RENO and fourth for Rovno. Then four terms for Double Chooz, Daya Bay, RENO and Bugey-4 \cite{declais} cross sections.
The minimization was made recurrently. At each step we have calculated $\chi^{2}_{k}$ and if the changing of individual nuclide spectrum shape leads to decreasing of $\chi^{2}$value we fix it in the database. The procedure was repeated until difference between neighbouring steps exceeds the chosen level $\epsilon$ until $|\chi^{2}_{k} - \chi^{2}_{k-1}| < \epsilon$. We used $\epsilon$ = 0.001.

\begin{figure*}[ht]
\begin{center}
\begin{minipage}{\textwidth}
\includegraphics[width=80mm]{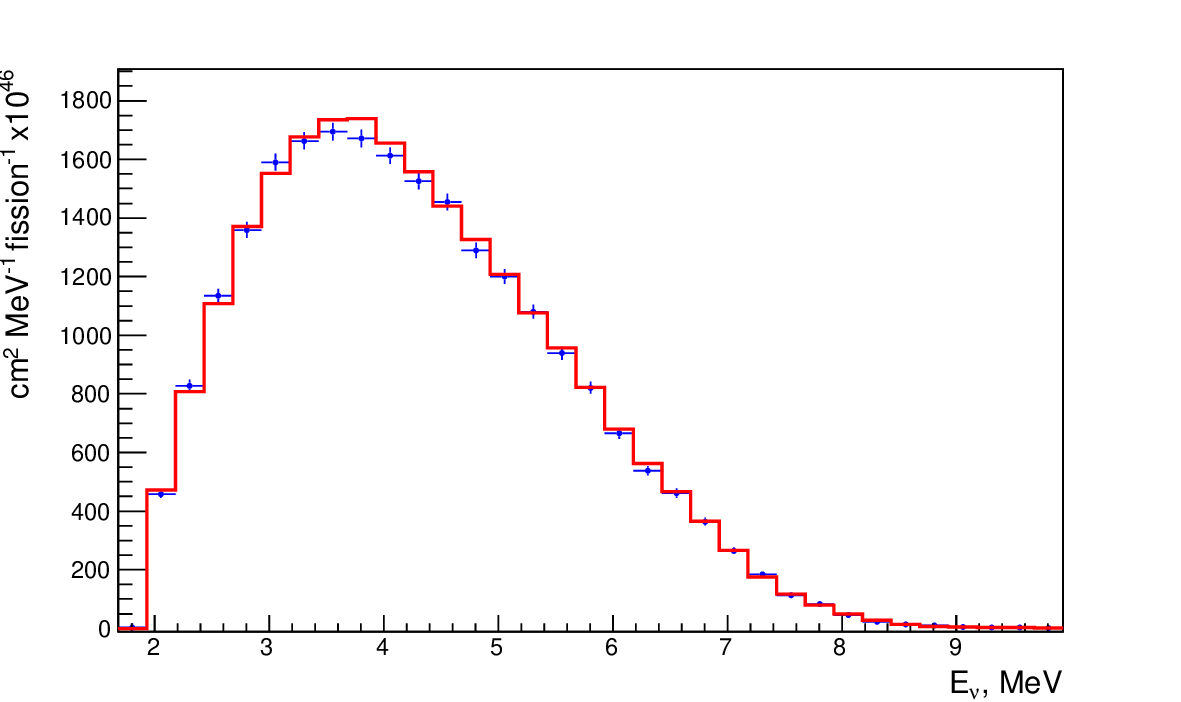}
\includegraphics[width=80mm]{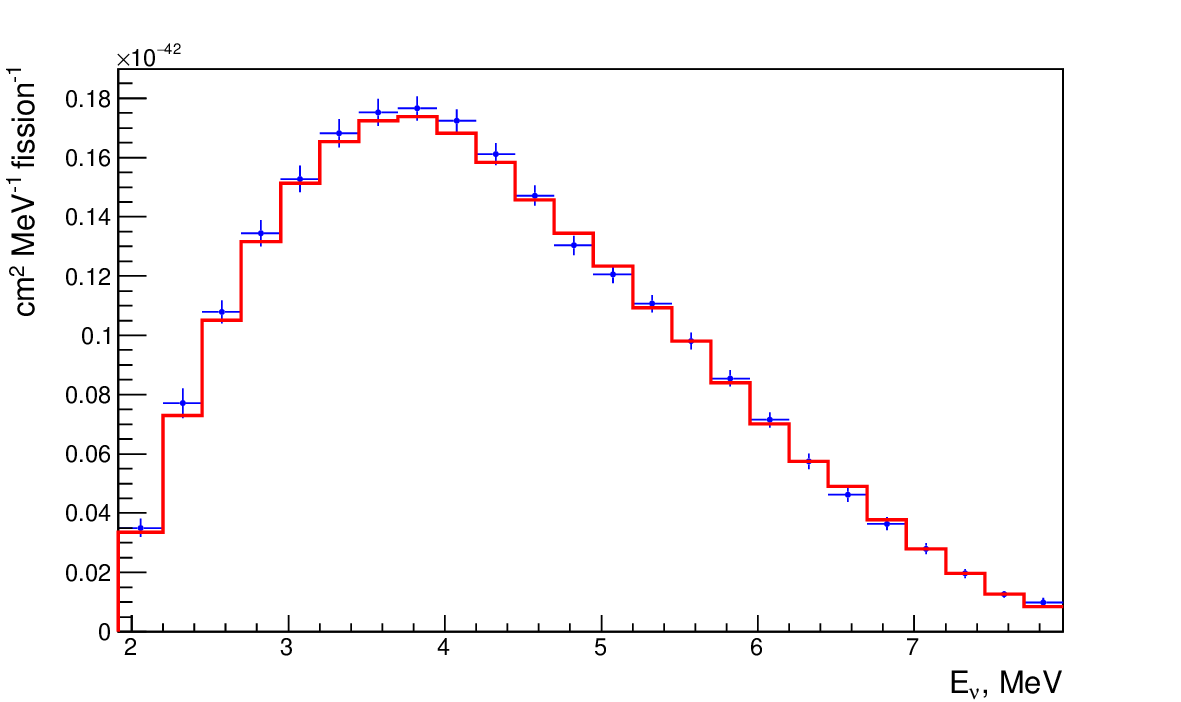}
\includegraphics[width=80mm]{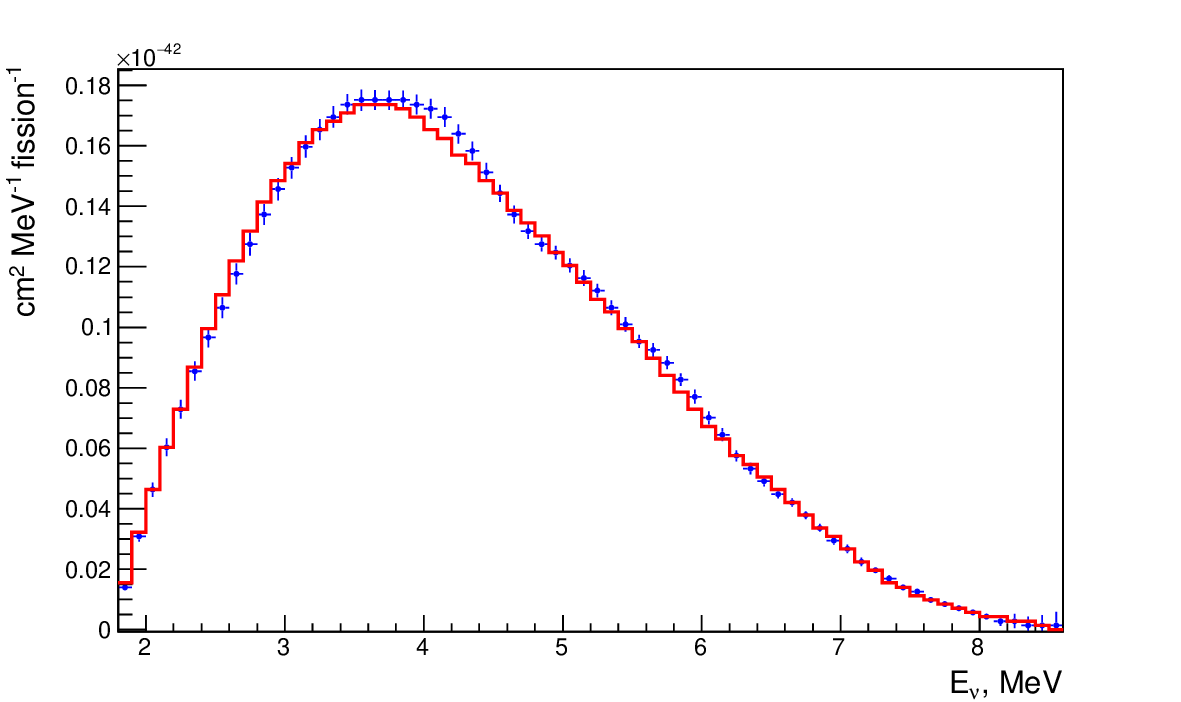}
\includegraphics[width=80mm]{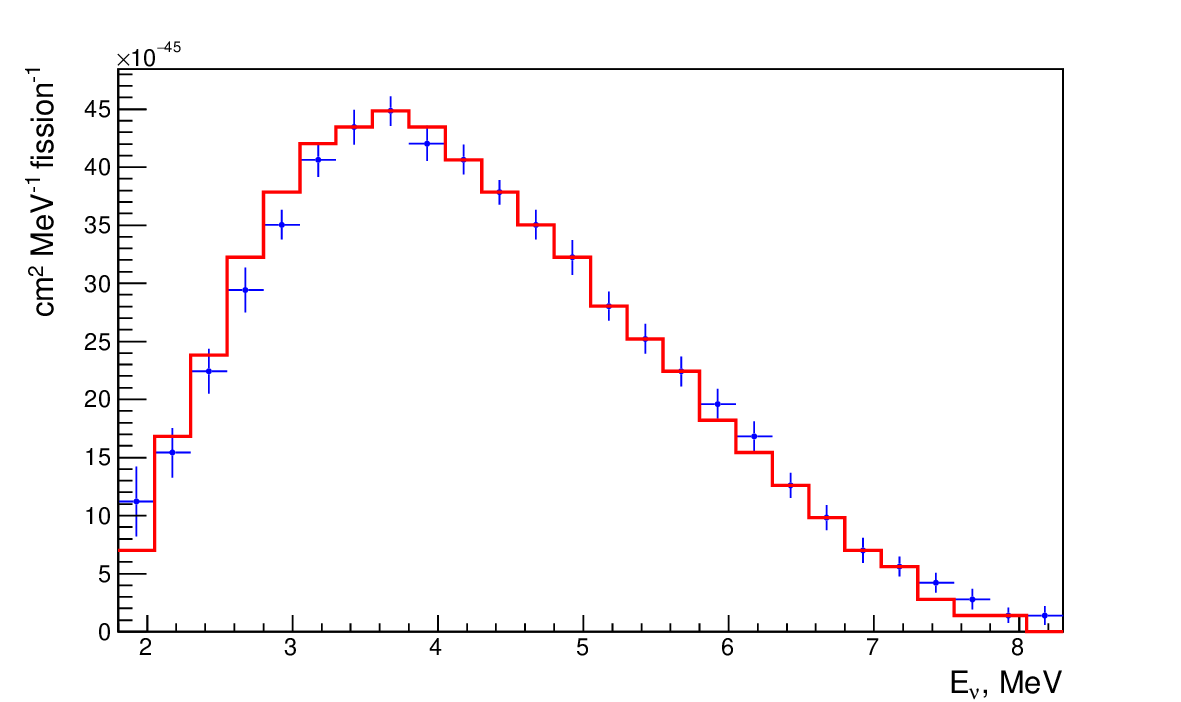}
\end{minipage}\hspace{2pc}
\end{center}
\caption{\label{figthree} Experimental antineutrino spectra weighted with IBD cross sections and our spectrum composition. Double Chooz (upper left panel in logarithmic scale), Daya Bay (upper right panel), RENO (down left panel) and PROSPECT (downright panel). Calculated spectra for individual core contents using INR spectra shown as red histogram.}
\end{figure*}

As a result of minimization the functional (\ref{chi2}) we got four fissile isotopes antineutrino spectra conformed all experimental reactor spectra. Measured and calculated antineutrino spectra weighted with IBD cross section shown at Fig. \ref{figthree}. Perfect similarity is seen for shown Double Chooz and Daya Bay spectra. Some roughs are seen for the RENO spectrum. And measured spectrum by PROSPECT where practically pure $^{235}$U spectrum exists also described with our spectrum. 

For a long time, the only one high precision accuracy measurement of IBD cross section was performed. The experiment at Bugey power plant used integral water detector registering only neutrons from reaction (1) was done \cite{declais} and measured IBD cross section in a reactor flux with accuracy 1.5\%. Recently this value was added with three more measurements Day Bay with accuracy 2\% \cite{an} and 1.2\% \cite{newdb} , RENO with accuracy 1.6\% \cite{yoon} and Double Chooz 1\% \cite{kerret}. The results of IBD cross sections measurements are presented in Table \ref{tabl:sect}. Here also one can find the prediction for cross section at pointed up core content using our spectra. 

To get cross section for $^{235}$U one can use simple relation between measured cross section and $^{235}$U one. 

\begin{equation}
\sigma_{f} = \sigma_{235}\cdot(1 + k),
\label{sigcor}
\end{equation}
where $\sigma_{f} -$ measured cross section, $\sigma_{235} -$ $^{235}$U cross section, $k -$ small negative correction due to core content. So, $^{235}$U cross section, should be a little bit larger than experimental one.

\begin{equation}
 k = f_{238}\cdot(\frac{\sigma_{238}}{\sigma_{235}} - 1)+f_{239}\cdot(\frac{\sigma_{239}}{\sigma_{235}} - 1)+f_{241}\cdot(\frac{\sigma_{241}}{\sigma_{235}} - 1) ,
\label{content}
\end{equation}
where $f_{i} –$ part of fissions for ith isotope, $\sigma_{i} –$ cross section for ith isotope ($i$ runs through 238, 239, 241). In the last column of the Table \ref{tabl:sect} $^{235}$U cross sections are presented that were taken from  appropriate experiment. 

\begin{table*}[ht]
%\multicolumn{2}
\caption{Cross sections of IBD reaction measured at different core content and calculated ones using calculated antineutrino spectra for fissile isotopes. Core content done in part of fissions and cross sections in cm$^2$/fission $\times10^{-43}$.}
\label{tabl:sect}
\centering
\vspace{2mm}
\begin{tabular}{ c | c | c | c | c | c | c | c }
\hline
\hline
Experiment & {Core content} & &&& $\sigma_{f}$ & $^{INR}\sigma_{f}$ & $\sigma_{235}$ \\
\hline
 & $^{235}$U & $^{238}$U & $^{239}$Pu & $^{241}$Pu &  &  &  \\
\hline
Double Chooz  & 0.520 & 0.087 & 0.333 & 0.06 & 5.71$\pm$0.06 & 5.820 & 5.76 \\
Daya Bay \cite{newdb} &0.561 & 0.076 & 0.307 & 0.056 & 5.84$\pm$0.07 & 5.804 & 5.98 \\
RENO & 0.571 & 0.073 & 0.300 & 0.056 & 5.852$\pm$0.094 &  5.801 & 5.92 \\
Bugey-4  & 0.538 & 0.078 & 0.328 & 0.056 & 5.752$\pm$0.081 & 5.782 & 5.84 \\
\hline
\hline
\end{tabular}	
\end{table*}

In Table \ref{tabl:bor1} one can see cross sections corresponding to calculated spectra from our fit. Averaged value through the last column of Table \ref{tabl:sect} is 5.86 what is practically the same as the section from our spectrum shown in Table \ref{tabl:bor1}. Also it coincides with the value 5.93 taken from fit of experimental cross sections and shown in last column of Table \ref{tabl:bor1}. 

If calculated spectra are correct, that means they are most close to really existing in Nature, the value of $\sigma_{235}$ from experimental measurement should be close to the value calculated through the spectrum. And values from different experiments be laying around the calculated one. But if spectra are incorrect all $\sigma_{235}$ from experimental measurements appear smaller than calculated one. 

We have fit experimental cross sections \cite{kerret}, \cite{an}, \cite{yoon}, \cite{declais} by the sum of parameters with weights equal to the core compositions of experiments. So, the mean values for isotope cross sections were obtained. 

\begin{equation}
 \sigma_{exp} = \sum_{i=235,238,239,24}  f_{i}\cdot par_{i} ,
\label{content}
\end{equation}
where $f_{i} -$ fission part for $i$th isotope, $par_{i} -$ cross section for $i$th isotope as parameter of fitting. Results of fitting the known four cross section measurements are presented in Table \ref{tabl:fitcs}.

\begin{table*}[ht] 
\caption{IBD reaction Cross sections for fissile isotopes from the fit of experimentally measured cross sections in cm$^2$/fission $\times10^{-43}$.}
\label{tabl:fitcs}
\centering
\vspace{2mm}
\begin{tabular}{ c | c | c | c | c | c }
\hline
\hline
Experiment & measured $\sigma_{f} $ & $^{235}$U & $^{238}$U & $^{239}$Pu & $^{241}$Pu \\
\hline
Double Chooz & 5.71$\pm$ 0.06 & 5.88 $\pm$ 0.6 & 9.38 $\pm$ 1.7 & 4.47 $\pm$ 0.8 & 5.81 $\pm$ 0.8 \\
Daya Bay \cite{an} & 5.91 $\pm$ 0.12 & 5.97 $\pm$ 0.8 & 10.29 $\pm$ 2.1 & 4.62 $\pm$ 0.8 & 6.42 $\pm$ 0.6 \\
Daya Bay \cite{newdb} & 5.84 $\pm$ 0.07 & 5.94 $\pm$ 0.6 & 9.85 $\pm$ 1.9 & 4.57 $\pm$ 0.7 & 6.36 $\pm$ 0.5 \\
RENO & 5.85 $\pm$ 0.09 & 5.94 $\pm$ 0.7 & 9.92 $\pm$ 1.6 & 4.58 $\pm$ 0.8 & 6.39 $\pm$ 0.5 \\
Bugey-4 & 5.75 $\pm$ 0.08 & 5.91 $\pm$ 0.8 & 9.58 $\pm$ 1.9 & 4.52 $\pm$ 0.6 & 6.11 $\pm$ 0.8 \\
\hline
\hline
\end{tabular}
\end{table*}

In the last column of the Table \ref{tabl:sect} one can find the $^{235}$U cross sections that were taken from the appropriate experiment. 
They can be compared with the last column of Table \ref{tabl:bor1} where mean value of {\it ith} cross section done.

\begin{table*}[ht] 
\caption{IBD reaction Cross sections for fissile isotopes antineutrino spectra and mean ones from the fit of experimentally measured cross sections in cm$^2$/fission $\times10^{-43}$.}
\label{tabl:bor1}
\centering
\vspace{2mm}
\begin{tabular}{ c | c | c }
\hline
\hline
Isotope & $\sigma_{f}$ & $^{fit}\sigma_{f}$  \\
\hline
$^{235}$U & 5.886 & 5.93 $\pm$ 0.6 \\
$^{238}$U & 10.93 & 9.74  $\pm$ 1.3 \\
$^{239}$Pu & 4.326 & 4.55  $\pm$ 0.7 \\
$^{241}$Pu & 6.656 & 6.25  $\pm$ 0.7 \\
\hline
\hline
\end{tabular}
\end{table*}

\begin{figure}[ht]
\begin{center}
\includegraphics[width=86mm]{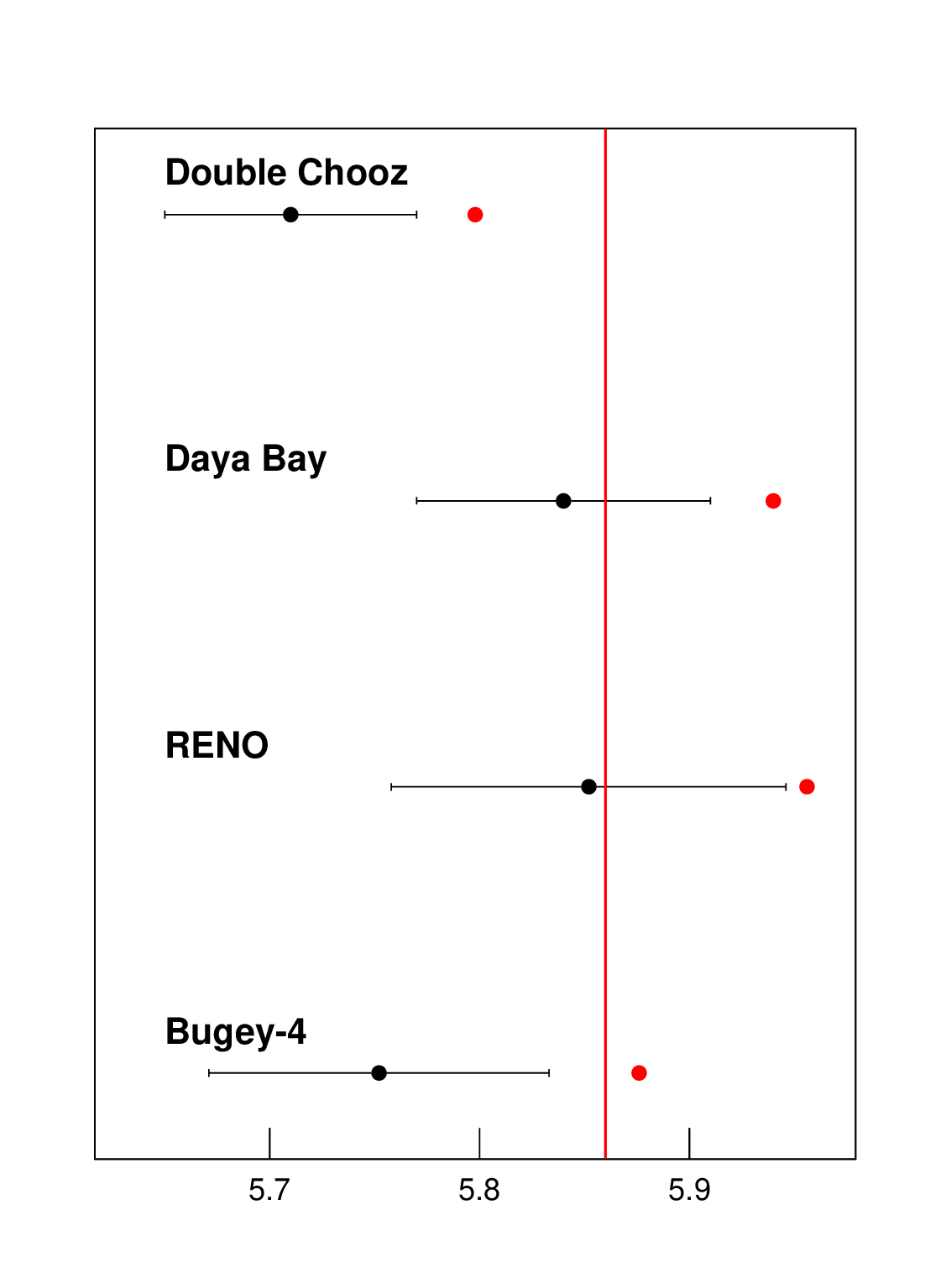}
\end{center}
\caption{\label{figfour} Experimental IBD cross sections (black dots with errors) and $^{235}$U spectrum cross sections (red dots) extracted from the experimental values. Red line marks the averaged $^{235}$U cross section over data from Table \ref{tabl:sect}.}
\end{figure}

Cross sections from the last measurements are shown at Fig. \ref{figfour}. Here also $^{235}$U cross sections from Table \ref{tabl:sect} shown in red dots. It is seen that the value of experimental cross section from Double Chooz experiment is the smallest one, but it should be the largest one due to core content. Its value makes $^{235}$U cross section mean value smaller than it could be.

\section{Conclusion}

The four antineutrino spectra $^{235}$U, $^{238}$U, $^{239}$Pu and $^{241}$Pu are performed. They were taken from the procedure of fitting experimental antineutrino spectra by the sum of calculated ones. Calculated spectra were taken by variation of branch probabilities for unknown fragments. Found spectra of fissile isotopes are presented in Table \ref{tabl:cnoflux}.

These spectra perfectly describe experimental IBD cross sections and their experimental spectra. Double Chooz cross section looks slightly smaller than it should be. Core content points up at larger value, this cross section should be the largest between all experimental sections. It is interesting that $^{235}$U cross section appeared smaller than $^{241}$Pu one. 

New beta-decay database for nuclides placed far from the line of beta-stability was developed.

We propose a new method of exploring short half-life nuclides by fitting measured antineutrino spectrum from nuclear reactor.

\vspace{2mm}
\section*{Acknowledgments}

We are grateful to Leonid B. Bezrukov for valuable discussions and useful remarks.

\vspace{2mm}
\section*{Conflict of interests}

The authors declare that they have no conflicts of interest.

\begin{table*}[ht] 
\caption{Antineutrino spectra of fissile isotopes in units $\bar{\nu}$ MeV$^{-1}$fission$^{-1}$.}
\label{tabl:cnoflux}
\centering
\vspace{2mm}
\begin{tabular}{ c | c | c | c | c }
\hline
\hline
$E_{\nu}$, MeV & $^{235}$U & $^{238}$U &$^{239}$Pu &$^{241}$Pu \\
\hline
1.5 & 1.823 & 2.062 & 1.583 & 1.813 \\
1.75 &  1.539 & 1.780 & 1.323 & 1.539 \\
2.0 & 1.209 & 1.458 & 1.048 & 1.245 \\
2.25 & 1.009 & 1.273 & 8.673$\cdot10^{-1}$ & 1.063\\
2.5 & 8.458$\cdot10^{-1}$ & 1.113 & 7.183$\cdot10^{-1}$ & 9.000$\cdot10^{-1}$\\
2.75 & 7.062$\cdot10^{-1}$ & 9.683$\cdot10^{-1}$ & 5.911$\cdot10^{-1}$ & 7.572$\cdot10^{-1}$\\
3.0 & 5.808$\cdot10^{-1}$ & 8.336$\cdot10^{-1}$ & 4.774$\cdot10^{-1}$ & 6.306$\cdot10^{-1}$\\
3.25 & 4.796$\cdot10^{-1}$ & 7.182$\cdot10^{-1}$ & 3.831$\cdot10^{-1}$ & 5.227$\cdot10^{-1}$\\
3.5 & 3.905$\cdot10^{-1}$ & 6.135$\cdot10^{-1}$ & 3.038$\cdot10^{-1}$ & 4.298$\cdot10^{-1}$\\
3.75 & 3.168$\cdot10^{-1}$ & 5.220$\cdot10^{-1}$ & 2.409$\cdot10^{-1}$ & 3.535$\cdot10^{-1}$\\
4.0 & 2.507$\cdot10^{-1}$ & 4.354$\cdot10^{-1}$ & 1.851$\cdot10^{-1}$ & 2.835$\cdot10^{-1}$\\
4.25 & 1.978$\cdot10^{-1}$ & 3.602$\cdot10^{-1}$ & 1.408$\cdot10^{-1}$ & 2.243$\cdot10^{-1}$\\
4.5 & 1.560$\cdot10^{-1}$ & 2.963$\cdot10^{-1}$ & 1.081$\cdot10^{-1}$ & 1.771$\cdot10^{-1}$\\
4.75 & 1.235$\cdot10^{-1}$ & 2.441$\cdot10^{-1}$ & 8.325$\cdot10^{-2}$ & 1.399$\cdot10^{-1}$\\
5.0 & 9.768$\cdot10^{-2}$ & 2.018$\cdot10^{-1}$ & 6.524$\cdot10^{-2}$ & 1.123$\cdot10^{-1}$\\
5.25 & 7.602$\cdot10^{-2}$ & 1.646$\cdot10^{-1}$ & 5.037$\cdot10^{-2}$ & 8.871$\cdot10^{-2}$\\
5.5 & 5.972$\cdot10^{-2}$ & 1.343$\cdot10^{-1}$ & 3.930$\cdot10^{-2}$ & 7.036$\cdot10^{-2}$\\
5.75 & 4.592$\cdot10^{-2}$ & 1.071$\cdot10^{-1}$ & 2.987$\cdot10^{-2}$ & 5.424$\cdot10^{-2}$\\
6.0 & 3.410$\cdot10^{-2}$ & 8.279$\cdot10^{-2}$ & 2.183$\cdot10^{-2}$ & 4.033$\cdot10^{-2}$\\
6.25 & 2.520$\cdot10^{-2}$ & 6.379$\cdot10^{-2}$ & 1.603$\cdot10^{-2}$ & 3.000$\cdot10^{-2}$\\
6.5 & 1.895$\cdot10^{-2}$ & 4.953$\cdot10^{-2}$ & 1.205$\cdot10^{-2}$ & 2.286$\cdot10^{-2}$\\
6.75 & 1.357$\cdot10^{-2}$ & 3.715$\cdot10^{-2}$ & 8.640$\cdot10^{-3}$ & 1.675$\cdot10^{-2}$\\
7.0 & 9.027$\cdot10^{-3}$ & 2.666$\cdot10^{-2}$ & 5.706$\cdot10^{-3}$ & 1.151$\cdot10^{-2}$\\
7.25 & 5.444$\cdot10^{-3}$ & 1.805$\cdot10^{-2}$ & 3.364$\cdot10^{-3}$ & 7.220$\cdot10^{-3}$\\
7.5 & 3.146$\cdot10^{-3}$ & 1.189$\cdot10^{-2}$ & 1.985$\cdot10^{-3}$ & 4.463$\cdot10^{-3}$\\
7.75 & 1.945$\cdot10^{-3}$ & 7.907$\cdot10^{-3}$ & 1.249$\cdot10^{-3}$ & 2.847$\cdot10^{-3}$\\
8.0 & 1.086$\cdot10^{-3}$ & 4.845$\cdot10^{-3}$ & 7.453$\cdot10^{-4}$ & 1.718$\cdot10^{-3}$\\
8.25 & 5.686$\cdot10^{-4}$ & 2.777$\cdot10^{-3}$ & 4.119$\cdot10^{-4}$ & 9.548$\cdot10^{-4}$\\
8.5 & 2.551$\cdot10^{-4}$ & 1.362$\cdot10^{-3}$ & 1.694$\cdot10^{-4}$ & 3.781$\cdot10^{-4}$\\
8.75 & 1.216$\cdot10^{-4}$ & 7.098$\cdot10^{-4}$ & 3.607$\cdot10^{-5}$ & 8.798$\cdot10^{-5}$\\
9.0 & 6.347$\cdot10^{-5}$ & 4.776$\cdot10^{-4}$ & 1.258$\cdot10^{-5}$ & 3.887$\cdot10^{-5}$\\
9.25 & 4.030$\cdot10^{-5}$ & 3.337$\cdot10^{-4}$ & 7.735$\cdot10^{-6}$ & 2.490$\cdot10^{-5}$\\
9.5 & 2.791$\cdot10^{-5}$ & 2.393$\cdot10^{-4}$ & 5.239$\cdot10^{-6}$ & 1.657$\cdot10^{-5}$\\
9.75 & 1.920$\cdot10^{-5}$ & 1.677$\cdot10^{-4}$ & 3.480$\cdot10^{-6}$ & 1.051$\cdot10^{-5}$\\
10.0 & 1.183$\cdot10^{-5}$ & 1.077$\cdot10^{-4}$ & 2.083$\cdot10^{-6}$ & 5.692$\cdot10^{-6}$\\
10.25 & 7.102$\cdot10^{-6}$ & 7.294$\cdot10^{-5}$ & 1.235$\cdot10^{-6}$ & 3.489$\cdot10^{-6}$\\
10.5 & 4.331$\cdot10^{-6}$ & 4.788$\cdot10^{-5}$ & 7.820$\cdot10^{-7}$ & 2.147$\cdot10^{-6}$\\
10.75 & 2.205$\cdot10^{-6}$ & 3.044$\cdot10^{-5}$ & 4.400$\cdot10^{-7}$ & 1.215$\cdot10^{-6}$\\
11.0 & 1.095$\cdot10^{-6}$ & 2.042$\cdot10^{-5}$ & 2.489$\cdot10^{-7}$ & 7.058$\cdot10^{-7}$\\
11.25 & 8.246$\cdot10^{-7}$ & 1.537$\cdot10^{-5}$ & 1.906$\cdot10^{-7}$ & 5.197$\cdot10^{-7}$\\
11.5 & 6.765$\cdot10^{-7}$ & 1.237$\cdot10^{-5}$ & 1.527$\cdot10^{-7}$ & 4.136$\cdot10^{-7}$\\
11.75 & 5.406$\cdot10^{-7}$ & 9.693$\cdot10^{-6}$ & 1.201$\cdot10^{-7}$ & 3.185$\cdot10^{-7}$\\
12.0 & 4.285$\cdot10^{-7}$ & 7.556$\cdot10^{-6}$ & 9.510$\cdot10^{-8}$ & 2.449$\cdot10^{-7}$\\
\hline
\hline
\end{tabular}	
\end{table*}

\end{document}